\documentstyle[12pt]{article}
\newcommand{\be}{\begin{eqnarray}}
\newcommand{\ee}{\end{eqnarray}}

\def\lsim{\mathrel{\rlap{\lower4pt\hbox{\hskip1pt$\sim$}}
    \raise1pt\hbox{$<$}}}               
\def\gsim{\mathrel{\rlap{\lower4pt\hbox{\hskip1pt$\sim$}}
    \raise1pt\hbox{$>$}}}               

\input{epsfig.sty}

\begin{document}

\Huge{\noindent{Istituto\\Nazionale\\Fisica\\Nucleare}}

\vspace{-3.9cm}

\Large{\rightline{Sezione SANIT\`{A}}}
\normalsize{}
\rightline{Istituto Superiore di Sanit\`{a}}
\rightline{Viale Regina Elena 299}
\rightline{I-00161 Roma, Italy}

\vspace{0.65cm}

\rightline{INFN-ISS 98/2}
\rightline{February 1998}

\vspace{1.5cm}

\begin{center}

\LARGE{Right-handed currents in rare exclusive\\ $B \to (K, K^*) \nu
\bar{\nu}$ decays}\\

\vspace{1cm}

\large{D. Melikhov$^{(a)}$, N. Nikitin$^{(a)}$ and S. Simula$^{(b)}$}\\

\vspace{0.5cm}

\normalsize{$^{(a)}$Nuclear Physics Institute, Moscow State University\\
Moscow, 119899, Russia\\ $^{(b)}$Istituto Nazionale di Fisica Nucleare,
Sezione Sanit\`{a}\\ Viale Regina Elena 299, I-00161 Roma, Italy}

\end{center}

\vspace{1cm}

\begin{abstract}

\noindent The effects of possible right-handed weak hadronic currents
in rare exclusive semileptonic decays $B \to (K, K^*) \nu \bar{\nu}$ are
investigated using a lattice-constrained dispersion quark model for the
calculation of the relevant mesonic form factors. The results obtained
for the branching ratios and the missing energy spectra are presented and
the sensitivity of various observables to long-distance physics is
investigated. It is shown that the asymmetry of transversely polarized
$K_T^*$ mesons as well as the $K / K_T^*$ production ratio are only
slightly sensitive to long-distance contributions and mostly governed by
the relative strength and phase of right-handed currents. In particular,
within the Standard Model the production of right-handed $K_T^*$ mesons
turns out to be largely suppressed with respect to left-handed ones,
thanks to the smallness of the final to initial meson mass ratio.
Therefore, the measurement of produced right-handed $K_T^*$ mesons in
rare $B \to K^* \nu \bar{\nu}$ decays offers a very interesting tool to
investigate right-handed weak hadronic currents. 

\end{abstract}

\vspace{0.5cm}


\newpage

\pagestyle{plain}

\indent Weak decays induced by Flavor Changing Neutral Currents ($FCNC$)
are widely recognized as a powerful tool to make stringent test of the
Standard Model ($SM$) as well as to probe possible New Physics ($NP$)
\cite{BBMR91}. Our understanding of $FCNC$ in terms of the $SM$ and its
possible extensions is expected to be improved by the foreseen advent of
new accelerators and $B$-factories, which will allow to investigate rare
$B$-meson decays induced by the $b \to s$ (and $b \to d$) transitions.
Besides many interesting processes, like, e.g., the $B_s - \bar{B}_s$
mixing and the rare $B$-meson decays induced by the $b \to s \gamma$ and
$b \to s \ell^+ \ell^-$ processes, the rare semileptonic decay $b \to s
\nu \bar{\nu}$ plays a peculiar role. Indeed, within the $SM$ the process
$b \to s \nu \bar{\nu}$ is governed by the following effective weak
Hamiltonian (cf. Refs. \cite{IL81,BB93,ALI96})
 \be
    \label{HeffSM}
    {\cal{H}}_{eff}^{(SM)}(b \to s \nu \bar{\nu}) & = & {G_F \over
    \sqrt{2}} {\alpha_{em} \over 2\pi \sin^2{\theta_W}} ~ V_{tb} V_{ts}^*
    ~ X(x_t) ~ O_L(b \to s \nu \bar{\nu}) \nonumber \\
    & & \equiv c_L^{(SM)} ~ O_L(b \to s \nu
    \bar{\nu})
 \ee
where $O_L(b \to s \nu \bar{\nu}) \equiv (\bar{s} \gamma_{\mu} (1 -
\gamma_5) b) ~ (\bar{\nu} \gamma^{\mu} (1 - \gamma_5) \nu)$. The
operator (\ref{HeffSM}) is obtained from $Z$-penguin and $W$-box diagrams
with a dominating top-quark intermediate state. In Eq. (\ref{HeffSM})
$G_F$ is the Fermi constant, $\alpha_{em}$ the fine structure constant,
$\theta_W$ the Weinberg angle, $V_{q q'}$ the Cabibbo-Kobayashi-Maskawa
($CKM$) matrix elements and $x_t \equiv (m_t / m_W)^2$; finally, the
function $X(x_t)$ is obtained after integrating out the heavy particles
and includes $O(\alpha_s)$ corrections (see Refs. \cite{IL81,BB93} for
its explicit expression).

\indent The appealing feature of Eq. (\ref{HeffSM}) relies in the
presence of a single operator governing the transition $b \to s \nu
\bar{\nu}$. In this way the main theoretical uncertainties are
concentrated in the value of only one Wilson coefficient,
$c_L^{(SM)}$\footnote{It turns out that the main uncertainty on
$c_L^{(SM)}$ is the uncertainty on the top-quark mass and on the product
$|V_{tb} V_{ts}^*|$ of $CKM$ matrix elements \cite{ALI96}, being the
radiative $QCD$ corrections substantially small \cite{BB93}.}. As it is
well known, in case of other processes, like those driven by the $b \to s
\gamma$ and $b \to s \ell^+ \ell^-$ transitions, several operators should
be included in the effective weak $SM$ Hamiltonian, so that the
corresponding set of Wilson coefficients (with their theoretical
uncertainties) act coherently in determining the values of many
observables, like the branching ratios, the differential decay rates,
lepton asymmetries, etc. Moreover, the $SM$ operator (\ref{HeffSM}) does
not contain long-distance contributions generated by four-quark
operators, which are usually present in the low-energy weak Hamiltonian
and affect both the $b \to s \ell^+ \ell^-$ and (to a much less extent)
the $b \to s \gamma$ processes.

\indent Under the only assumption of purely left-handed neutrinos
(possible neutrino mass effects are expected to be negligible
\cite{GLN96}) $NP$ effects in the $b \to s \nu \bar{\nu}$ transitions can
be a modification of the $SM$ value of the coefficient $c_L$ and/or the
introduction of a new right-handed ($RH$) operator, viz. 
 \be
   \label{Heff}
   {\cal{H}}_{eff}(b \to s \nu \bar{\nu}) = c_L ~ O_L(b \to s \nu
   \bar{\nu}) ~ + ~ c_R ~ O_R(b \to s \nu \bar{\nu})
 \ee
where $O_R(b \to s \nu \bar{\nu}) \equiv (\bar{s} \gamma_{\mu} (1 +
\gamma_5) b) ~ (\bar{\nu} \gamma^{\mu} (1 - \gamma_5) \nu)$ and the values
of the coefficients $c_L$ and $c_R$ depend on the specific $NP$ model.
In what follows, we will make use of two parameters, $\epsilon$ and
$\eta$, defined as
 \be
    \label{epseta}
    \epsilon^2 \equiv {|c_L|^2 + |c_R|^2 \over |c_L^{(SM)}|^2},
     ~~~~~~~~~~~~~~~~
    \eta \equiv - {\mbox{Re}(c_L c_R^*) \over |c_L|^2 + |c_R|^2}
 \ee
which are clearly connected to the relative strength and phase of $RH$
currents with respect to left-handed ($LH$) ones.

\indent From the theoretical point of view the inclusive $B \to X_s \nu
\bar{\nu}$ decay is a particularly clean process for investigating
possible $NP$ effects, because the non-perturbative $1 / m_b^2$
corrections to the free-quark result are known to be small \cite{FLS96}.
This is valid not only for the branching ratio, but also for the
differential decay rate, except for the regions close to the kinematical
end-point, where the spectrum has to be smeared out to get reliable
results. The free-quark prediction for the missing-energy spectrum of the
$B \to X_s \nu \bar{\nu}$ decay can be read off from, e.g., Ref.
\cite{GLN96}. While the differential decay rate is directly proportional
to $\epsilon^2$, its dependence upon $\eta$ is due to interference
effects between the $LH$ and $RH$ currents. However, since the final to
initial quark mass ratio, $m_s / m_b$, is quite small, the shape of the
missing energy spectrum is only slightly affected by the value of $\eta$
(cf. Ref. \cite{GLN96}). Therefore, a full determination of the
coefficients $c_L$ and $c_R$ requires at least to consider also
decay processes other than the inclusive $B \to X_s \nu \bar{\nu}$ one.
It is the aim of this letter to show that the effects of possible $RH$
currents can be investigated in rare exclusive semileptonic decays $B \to
(K, K^*) \nu \bar{\nu}$ and to this end a lattice-constrained dispersion
quark model, recently developed in  Ref. \cite{MNS}, is used to evaluate
the relevant mesonic form factors. It will be shown that the asymmetry of
transversely polarized $K_T^*$ mesons as well as the $K / K_T^*$
production ratio are only slightly affected by the model dependence of
the form factors and remarkably sensitive both to $\epsilon$ and $\eta$,
i.e. to the relative strength and phase of $RH$ currents. In particular,
within the $SM$ the production of $RH$ $K_T^*$ mesons turns out to be
largely suppressed with respect to $LH$ ones, thanks to the smallness of
the final to initial meson mass ratio and, therefore, the measurement of
produced $RH$ $K_T^*$ mesons in rare $B \to K^* \nu \bar{\nu}$ decays
offers a very interesting tool to investigate $RH$ weak hadronic currents.

\indent To begin with, let us denote by $P_B$ and $P_{K(K^*)}$ the
four-momentum of the initial and final mesons and define $q = P_B -
P_{K(K^*)}$ as the four-momentum of the $\nu \bar{\nu}$ pair and $x
\equiv E_{miss} / M_B$ the missing energy fraction, which is related to
the squared four-momentum transfer $q^2$ by: $q^2 = M_B^2 ~ [ 2x - 1 +
r_{K(K^*)}^2]$, where $r_{K(K^*)} \equiv M_{K(K^*)} / M_B$ with $M_B$ and
$M_{K(K^*)}$ being the initial and final meson masses. The missing energy
spectrum for the decay $B \to K \nu \bar{\nu}$ can be written as (cf.
Refs. \cite{MNS,COL97})
 \be
    \label{BtoK}
    {dBr \over dx}(B \to K \nu \bar{\nu}) = 3 Br_0 ~ \left| {c_L + c_R
    \over c_L^{(SM)}} \right|^2 ~ \left[ (1 - x)^2 - r_K^2 \right]^{3/2}
    ~ |F_1(q^2)|^2
 \ee
where the factor $3$ arises from the sum over the three neutrino
generations and
 \be
    \label{Br0}
    Br_0 \equiv |c_L^{(SM)}|^2 {M_B^5 \tau_B \over 6 \pi^3} = {G_F^2 M_B^5
    \over (4 \pi)^3} {4 \alpha_{em}^2 \over 3 \pi^2 sin^4(\theta_W)} ~
    X^2(x_t) ~ |V_{tb} V_{ts}^*|^2 ~ \tau_B
 \ee
with $\tau_B$ being the $B$-meson lifetime. In Eq. (\ref{BtoK}) the
mesonic form factor $F_1(q^2)$ is obtained from the covariant
decomposition of the hadronic transition $B \to K$ driven by the vector
current $\bar{s} \gamma_{\mu} b$, viz.
 \be
    \label{F1}
    \langle K | \bar{s} \gamma_{\mu} b | B \rangle = (P_B + P_K)_{\mu}
    ~ F_1(q^2) ~ + ~ q_{\mu} ~ [F_0(q^2) - F_1(q^2)] ~ {M_B^2 - M_K^2
    \over q^2}
 \ee
with $F_0(q^2 = 0) = F_1(q^2 = 0)$. As for the $B \to K_h^* \nu \bar{\nu}$
decay, the missing energy spectrum corresponding to a definite
polarization $h$ ($= 0, \pm1$) of the final $K^*$ meson is given by
\cite{COL97}
 \be
    \label{BtoK*L}
    {dBr \over dx}(B \to K_{h = 0}^* \nu \bar{\nu}) & = & {3 \over 4} Br_0
    ~ \left| {c_L - c_R \over c_L^{(SM)}} \right|^2 ~ {1 \over r_{K^*}^2
    (1 + r_{K^*})^2} ~ \sqrt{(1 - x)^2 - r_{K^*}^2} \cdot \nonumber \\ 
    & & \left| (1 + r_{K^*})^2 (1 - x - r_{K^*}^2) ~ A_1(q^2) - 2 [(1 -
    x)^2 - r_{K^*}^2] ~ A_2(q^2) \right|^2
 \ee
 \be
    \label{BtoK*T}
    {dBr \over dx}(B \to K_{h = \pm 1}^* \nu \bar{\nu}) & = & {3 \over 4}
    Br_0 ~ \sqrt{(1 - x)^2 - r_{K^*}^2} ~ {2x - 1 + r_{K^*}^2 \over (1 +
    r_{K^*})^2} ~ \left| 2 {c_L + c_R \over c_L^{(SM)}} \cdot \right.
    \nonumber \\ 
    & & \left. \sqrt{(1 - x)^2 - r_{K^*}^2} ~ V(q^2) \mp {c_L - c_R \over
    c_L^{(SM)}} ~ (1 + r_{K^*})^2 ~ A_1(q^2) \right|^2
 \ee
where the mesonic form factors $V(q^2)$, $A_1(q^2)$ and $A_2(q^2)$ appear
in the covariant decomposition of the hadronic matrix elements of the $B
\to K^*$ transition generated by the $V - A$ current $\bar{s}
\gamma_{\mu} (1 - \gamma_5) b$, viz.
 \be
    \label{VA1A2}
    \langle K_h^* | \bar{s} \gamma_{\mu} (1 - \gamma_5) b | B \rangle &
    = & \epsilon_{\mu \nu \alpha \beta} e^{*\nu}(h) P_B^{\alpha}
    P_{K^*}^{\beta} {2 V(q^2) \over M_B + M_{K^*}} - i \left\{
    e_{\mu}^*(h) (M_B + M_{K^*}) A_1(q^2) - \right. \nonumber \\
    & & \left. [e^*(h) \cdot q] (P_B + P_{K^*})_{\mu} {A_2(q^2) \over M_B
    + M_{K^*}} - \right. \nonumber \\
    & & \left. [e^*(h) \cdot q] q_{\mu} {2 M_{K^*} \over q^2} [A_3(q^2)
    - A_0(q^2)] \right\}
 \ee
where $e(h)$ is the polarization four-vector of the $K^*$-meson and
$A_3(q^2) \equiv [(M_B + M_{K^*}) A_1(q^2) - (M_B - M_{K^*}) A_2(q^2)] /
2M_{K^*}$.

\indent In the whole accessible kinematical decay region we have
calculated the relevant mesonic form factors $F_1(q^2)$, $V(q^2)$,
$A_1(q^2)$ and $A_2(q^2)$, appearing in Eqs. (\ref{BtoK}) and
(\ref{BtoK*L}-\ref{BtoK*T}), adopting a dispersion formulation of the
relativistic quark model \cite{MEL96}. For the explicit evaluation of the
form factors one needs to specify the quark model parameters such as the
constituent quark masses and the meson wave functions. In Ref.
\cite{MNS97} we performed calculations of the mesonic form factors
adopting different model wave functions, in particular: the simple
Gaussian ans\"atz of the $ISGW2$ model \cite{ISGW} and the variational
solution \cite{CAR94} of the effective $q \bar{q}$ semi-relativistic
Hamiltonian of Godfrey and Isgur ($GI$) \cite{GI85}. These two models
differ both in the shape of the meson wave functions, particularly at
high internal momenta, and in the values of the quark masses (see Ref.
\cite{MNS97} for details). The results of our calculations showed that
the mesonic form factors for heavy-to-light transitions are sensitive both
to the high-momentum tail of the meson wave function and to the values
adopted for the quark masses (see also Ref. \cite{SIM96}). In order to
obtain more reliable predictions for the form factors, we have required
\cite{MNS} the quark model parameters to be adjusted in such a way that
the calculated form factors at high $q^2$ are compatible with recent
available lattice $QCD$ results \cite{APE,UKQCD}. We found that the best
agreement with the high-$q^2$ lattice data can be obtained adopting the
quark masses and wave functions of the $GI$ model with a switched-off
one-gluon exchange, which will be denoted hereafter to as the $GI - OGE$
quark model.

\indent Recently, in the whole range of accessible values of $q^2$ a
lattice-constrained parametrization for the $B\to K^*$ form factors has
been developed \cite{LAT97}, based on the Stech's parametrization of the
form factors obtained within the constituent quark picture \cite{STE95},
on the Heavy Quark Symmetry ($HQS$) scaling relations near $q^2 = M_B^2 -
M_{K(K^*)}^2$ and on a single-pole behavior of $A_1(q^2)$ suggested by the
heavy-quark mass dependence at $q^2 = 0$ expected from the $QCD$ sum
rules (see Ref. \cite{LAT97} for details). The parameters of the
single-pole fit to the form factor $A_1(q^2)$ were found from the
least-$\chi^2$ fit to the lattice $QCD$ simulations \cite{APE,UKQCD} in a
limited region at high values of $q^2$. Such a parametrization, though
still phenomenological, is also consistent with the dispersive bounds of
Ref. \cite{LEL96} and therefore it obeys all known theoretical
constraints. The comparison of the mesonic form factors relevant in rare
$B \to (K, K^*) \nu \bar{\nu}$ decays, obtained in our $GI - OGE$
relativistic quark model, with the parametrization of Ref. \cite{LAT97}
is shown in Fig. 1. It can clearly be seen that the two sets of form
factors agree each other within $\simeq 10 \%$, except near the
zero-recoil point. We want to stress that both sets of form factors
satisfy all known rigorous theoretical constraints in the whole
kinematical accessible region. Therefore, we expect that the difference
between the two sets of form factors provide a typical present-day
theoretical uncertainty of our knowledge of long-distance effects in the
mesonic channels. Consequently, in order to investigate the sensitivity
of the missing-energy spectra and branching ratios of rare $B \to (K,
K^*) \nu \bar{\nu}$ decays to the specific $q^2$-behavior of the relevant
form factors, we have run calculations of Eqs. (\ref{BtoK}) and
(\ref{BtoK*L}-\ref{BtoK*T}) adopting the two sets of mesonic form factors
shown in Fig. 1. We have also adopted the following values: $\tau_B =
(1.57 \pm 0.04) ~ ps$ \cite{BHP96}, $M_B = 5.279 ~ GeV$ \cite{PDG96},
$sin(\theta_W) = 0.2315$ \cite{PDG96} and $|V_{tb} V_{ts}^*| = 0.038 \pm
0.005$ ($0.041 \pm 0.005$) in case of our $GI - OGE$ form factors
(parametrization of Ref. \cite{LAT97})\footnote{The values adopted for
$|V_{tb} V_{ts}^*|$ for the two sets of form factors have been determined
in Ref. \cite{MNS} by the request of reproducing the $CLEO$ result
\cite{CLEO} on the rare exclusive $B \to K^* \gamma$ decay within the
$SM$ basis.}. Finally, at $m_t = 176 ~ GeV$ one gets $X(x_t) = 2.02$,
yielding $Br_0 = (5.3 \pm 1.4) \cdot 10^{-4}$ (see Eq. (\ref{Br0})),
which implies a present-day theoretical uncertainty of $\sim 25 \%$ in
rare $b \to s \nu \bar{\nu}$ decay rates.

\indent The present experimental upper bound on the inclusive branching
ratio $Br(B \to X_s \nu \bar{\nu})$, determined by the $ALEPH$
collaboration ($Br(B \to X_s \nu \bar{\nu}) < 7.7 \cdot 10^{-4}$
\cite{ALEPH}), turns out to be about an order of magnitude larger than the
typical $SM$ prediction ($Br_{(SM)}(B \to X_s \nu \bar{\nu}) \simeq 4
\div 5 \cdot 10^{-5}$ \cite{BB93,COL97}). Thus, till now the inclusive $B
\to X_s \nu \bar{\nu}$ decay constrains weakly the range of values of the
relative strength factor $\epsilon^2$, namely: $\epsilon^2 \lsim 15 \cdot
Br(B \to X_s \nu \bar{\nu}) / (7.7 \cdot 10^{-4})$ and, in general, the
admixture of possible $RH$ currents in $b \to s$ transitions is not yet
constrained too much, leaving the possibility of a sizable strength with
unknown relative phase (see, e.g., \cite{RIZ98}). Since the present $SM$ 
uncertainty on rare $b \to s \nu \bar{\nu}$ decy rates is about $25 \%$,
we have simply considered a relative strength factor $\epsilon^2$ equal
to $1.25$ (i.e., a $25 \%$ enhancement with respect to the $SM$ value
$\epsilon_{(SM)}^2 = 1$) and varied the parameter $\eta$ in its allowable
range. Note that the value $\epsilon^2 = 1$ can be realized not only
within the $SM$ framework, but possibly also in $NP$ scenarios; however,
for sake of simplicity, in what follows we will assume $c_L =
c_L^{(SM)}$. Our results obtained for the branching ratios $Br(B \to K
\nu \bar{\nu})$, $Br(B \to K_L^* \nu \bar{\nu})$ and $Br(B \to K_T^* \nu
\bar{\nu})$ (where $K_L^*$ and $K_T^*$ stand for longitudinally and
transversely polarized $K^*$-mesons, respectively), are reported in Fig.
2(a) and clearly show that the production of longitudinally polarized
$K_L^*$-mesons is remarkably sensitive to the detailed $q^2$-behavior of
the mesonic form factors, while our predictions for both $K_T^*$- and
$K$-meson production are only slightly model-dependent. Note that,
contrary to what happens in case of the inclusive $B \to X_s \nu
\bar{\nu}$ process, the exclusive branching ratios $Br[B \to (K, K^*) \nu
\bar{\nu}]$ are much more sensitive to the value of the parameter $\eta$,
i.e. to the relative phase of $RH$ currents with respect to $LH$ ones,
even when a quite small enhancement factor $\epsilon^2 - 1$ is
considered. Our predictions for the ratio $R_{K / K_T^*}$ of produced
$K$- to $K_T^*$-mesons as well as the transverse asymmetry $A_T$, defined
as
 \be
    \label{ratios}
    R_{K / K_T^*} \equiv {Br(B \to K \nu \bar{\nu}) \over Br(B \to K_{h =
    -1}^* \nu \bar{\nu}) + Br(B \to K_{h = +1}^* \nu \bar{\nu})}
    \nonumber \\
    A_T \equiv {Br(B \to K_{h = -1}^* \nu \bar{\nu}) - Br(B \to K_{h =
    +1}^* \nu \bar{\nu}) \over Br(B \to K_{h = -1}^* \nu \bar{\nu}) +
    Br(B \to K_{h = +1}^* \nu \bar{\nu})}
 \ee
are shown in Fig. 2(b). It can be seen that the transverse asymmetry
$A_T$ is only marginally affected by the model-dependence of the mesonic
form factors, while the ratio $R_{K / K_T^*}$ is more sensitive to their
specific $q^2$-behavior, particularly at negative values of $\eta$. It
should be pointed out however that both $A_T$ and (to a much larger
extent) $R_{K / K_T^*}$ are remarkably affected by the value of $\eta$.
The advantage of such a large sensitivity should be taken into account
when comparing inclusive versus exclusive decay modes, the latter being
affected by the general problem of the model dependence of the form
factors.

\indent It can be easily checked (starting from Eqs. (\ref{BtoK}) and
(\ref{BtoK*L}-\ref{BtoK*T})) that the ratio $R_{K / K_T^*}$ is
independent of $\epsilon$, while the asymmetry $A_T$ depends both on
$\eta$ and $\epsilon$\footnote{Note also that both $R_{K / K_T^*}$ and
$A_T$ are clearly independent of $|V_{tb} V_{ts}^*|$ and the top-quark
mass.}. Our predictions for the $SM$ values of $R_{K / K_T^*}$ and $A_T$
are $0.76 \pm 0.04$ (see Fig. 2(b) at $\eta = 0$) and $0.93 \pm 0.02$,
respectively, where the quoted uncertainties correspond to the variation
obtained using the two sets of mesonic form factors adopted in this work.
The $SM$ value of $A_T$ indicates a dominance of produced $LH$ ($h = -1$)
$K_T^*$-mesons with respect to $RH$ ($h = +1$) ones. This fact is clearly
illustrated in Fig. 3, where our predictions for the missing-energy
spectra of longitudinally and transversely polarized $K^*$-mesons,
obtained within the $SM$ framework, are reported. The main outcome can be
summarised as follows: i) the energy spectrum of longitudinally polarized
$K_L^*$-mesons is largely affected by the model-dependence of the mesonic
form factors, while the opposite feature is exhibited by the production
of transversely polarized $K_T^*$-mesons, and ii) $RH$ $K_T^*$-mesons are
less abundant than $LH$ ones within the $SM$ in a wide range of values
of $x$, except near the zero-recoil point (corresponding to $x = x_{max}
= 1 - r_{K^*}$). We have reached the same conclusions also after having
carried out the calculations of the transverse asymmetry $A_T$ using
other sets of form factors, like those obtained within the two $QCD$ sum
rule versions of Refs. \cite{COL97} and \cite{ALIEV97}, or the form
factors fulfilling the $HQS$ relations at leading-order in the inverse
heavy-quark mass (see Refs. \cite{IW,MNS97}). As for the latter case, it
is well known that in the heavy-quark limit ($HQL$) all the relevant form
factors are related to a single universal function, the Isgur-Wise form
factor. Moreover, in the $HQL$ the (differential) transverse asymmetry
$A_T(x)$ is independent of the Isgur-Wise function and within the $SM$ it
simply reduces to a kinematical function, viz:
 \be
    \label{asymmetry}
    A_T(x) \equiv {dBr(B \to K_{h = -1}^* \nu \bar{\nu}) / dx - dBr(B \to
    K_{h = +1}^* \nu \bar{\nu}) / dx \over dBr(B \to K_{h = -1}^* \nu
    \bar{\nu}) / dx + dBr(B \to K_{h = +1}^* \nu \bar{\nu}) / dx}
    \to_{HQL} {\sqrt{\omega^2 - 1} \over \omega}
 \ee
where $\omega$ is dot product of the initial and final meson
four-velocities. Therefore, starting from the zero-recoil point ($\omega
= 1$) where $A_T(x) = 0$, the transverse asymmetry in the $SM$ rapidly
increases up to its maximum value $(1 - r_{K^*}^2) / (1 + r_{K^*}^2)$,
reached at the maximum-recoil point $\omega_{max} = (1 + r_{K^*}^2) / 2
r_{K^*}$ (corresponding to $x = x_{min} = (1 - r_{K^*}^2) / 2$). Since
$r_{K^*}^2 \simeq 0.03$ one has $A_T(x) \sim 0.9$ in a wide range of
values of $x$. Though the final $K^*$-mesons are far from being
considered as heavy daughters, approximate $HQS$ relations among the form
factors of the $B \to K^*$ transition have been shown to hold within
$\simeq 20\%$ accuracy \cite{MNS,MNS97}), so that the dominance of $LH$
produced $K_T^*$ mesons within the $SM$ holds not only in the $HQL$, but
also in case of finite quark masses. To sum up, the measurement of
produced $RH$ $K_T^*$-mesons in rare $B \to K^* \nu \bar{\nu}$ decays
could offer a clear signature of possible $RH$ weak hadronic currents.
Finally, we have collected in Fig. 4 our predictions for the shape of the
missing-energy spectra of the $B \to K_{h = \pm 1}^* \nu \bar{\nu}$
decay, obtained for various values of $\epsilon$ at fixed value of $\eta$
(see Fig. 4(a)) as well as for various values of $\eta$ at fixed
$\epsilon$ (see Fig. 4(b)). Note in particular that the production of
$LH$ $K_T^*$-mesons is almost independent of the value of $\epsilon$,
while the differential rate for $RH$ ones is approximately proportional
to $(\epsilon^2 - 1)$.

\indent In conclusion, the missing-energy spectra and branching ratios
of rare exclusive semileptonic $B \to (K, K^*) \nu \bar{\nu}$ decays have
been investigated adopting a lattice-constrained dispersion quark model
for the calculation of the relevant mesonic form factors. The effects of
possible right-handed weak hadronic current have been considered and the
sensitivity of the branching ratios and the missing energy spectra to
long-distance physics has been investigated. It has been shown that the
asymmetry of transversely polarized $K_T^*$ mesons as well as the $K /
K_T^*$ production ratio are only slightly sensitive to long-distance
contributions and mostly governed by the relative strength and phase of
right-handed currents. In particular, within the Standard Model the
production of right-handed $K_T^*$ mesons turns out to be largely
suppressed with respect to left-handed ones, thanks to the smallness of
the final to initial meson mass ratio. Therefore, the measurement of
produced right-handed $K_T^*$ mesons in rare $B \to K^* \nu \bar{\nu}$
decays offers a very interesting tool to investigate right-handed weak
hadronic currents and, despite the general problem of the model
dependence of the hadronic form factors, the exclusive decay modes $B \to
(K, K^*) \nu \bar{\nu}$ turn out to be more sensitive to the effects of
right-handed currents with respect to the inclusive $B \to X_s \nu
\bar{\nu}$ process.

\newpage

\begin{figure}[htb]

\epsfxsize=10cm \epsfig{file=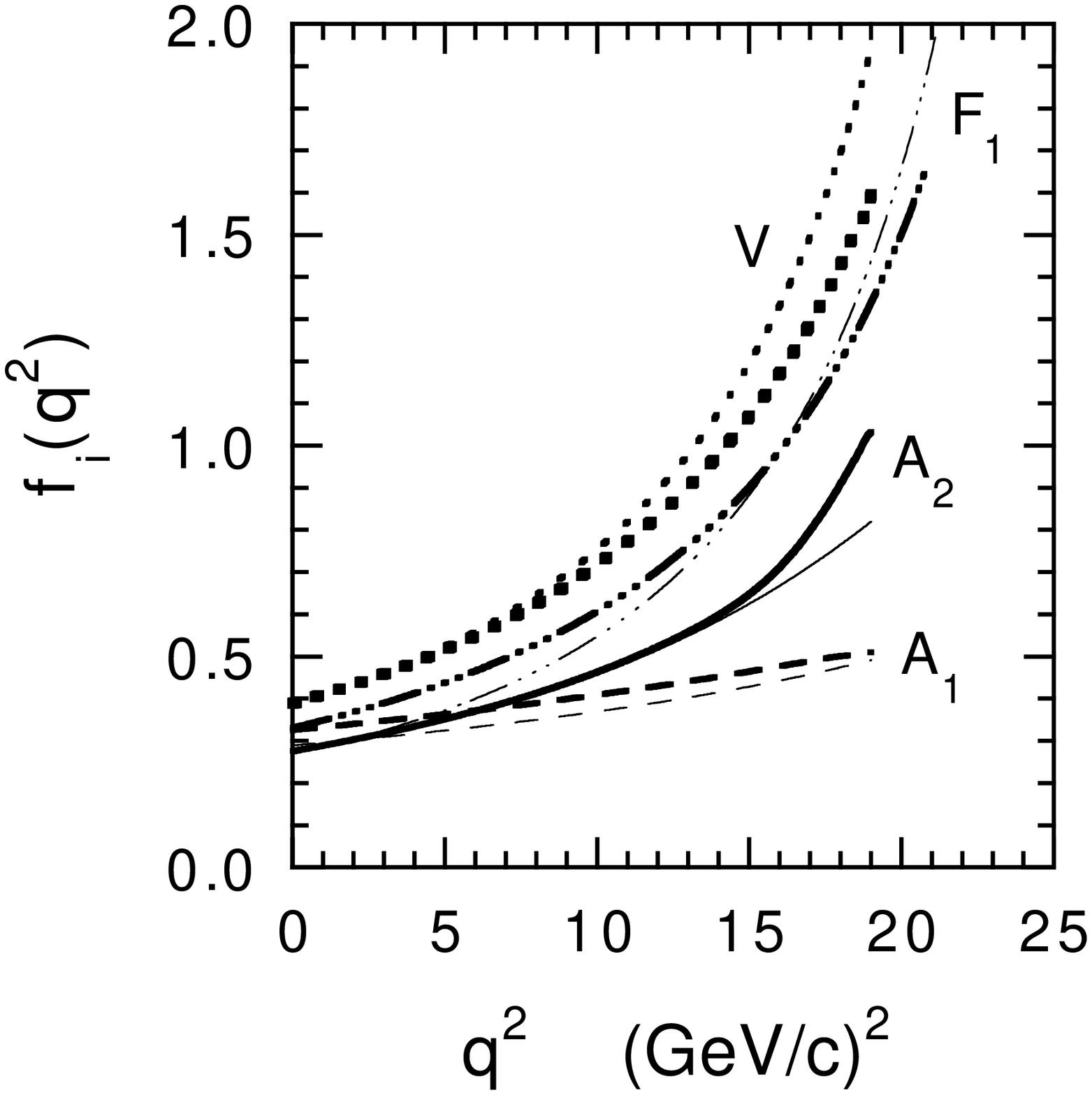}

\vskip -13cm

\parbox{10cm} \ $~~~~~~~~$ \ \parbox{6cm}{{\small \noindent
{\bf Figure 1}. The mesonic form factors $F_1(q^2)$, $V(q^2)$,
$A_1(q^2)$ and $A_2(q^2)$ relevant in rare $B \to (K, K^*) \nu \bar{\nu}$
decays as a function of the squared four-momentum $q^2$ of the $\nu
\bar{\nu}$ pair. Thin lines represent the parametrization of Ref.
\cite{LAT97}, while the thick lines correspond to the results of our
lattice-constrained dispersion quark picture, based on the $GI - OGE$
model (see text and Ref. \cite{MNS}).}}

\end{figure}

\vspace{2cm}

\begin{figure}[htb]

\epsfxsize=18cm \epsfig{file=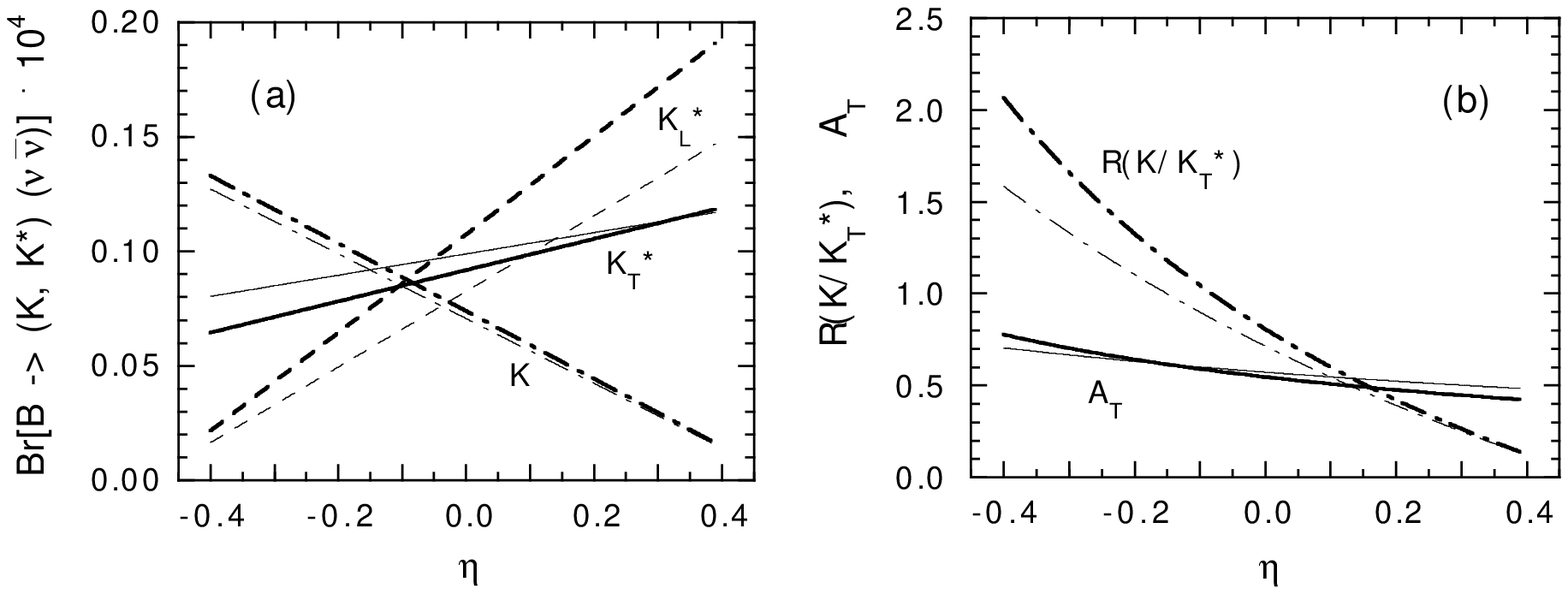}

\vspace{-18cm}

{\bf Figure 2}. {\small (a) Branching ratios of the rare exclusive 
processes $B \to K \nu \bar{\nu}$ (dot-long dashed lines), $B \to 
K_L^* \nu \bar{\nu}$ (dashed lines) and $B \to K_T^* \nu \bar{\nu}$ 
(solid lines), where $K_L^*$ ($K_T^*$) stands for longitudinally 
(transversely) polarized $K^*$-mesons, versus the parameter $\eta$ 
at $\epsilon^2 = 1.25$ (see Eq. (\ref{epseta})). The thin and thick 
lines have the same meaning as in Fig. 1. (b) The same as in (a), but 
for the ratio $R_{K / K_T^*}$ and the transverse asymmetry $A_T$ (see 
Eq. (\ref{ratios})).}

\end{figure}

\newpage

\begin{figure}[htb]

\epsfxsize=10cm \epsfig{file=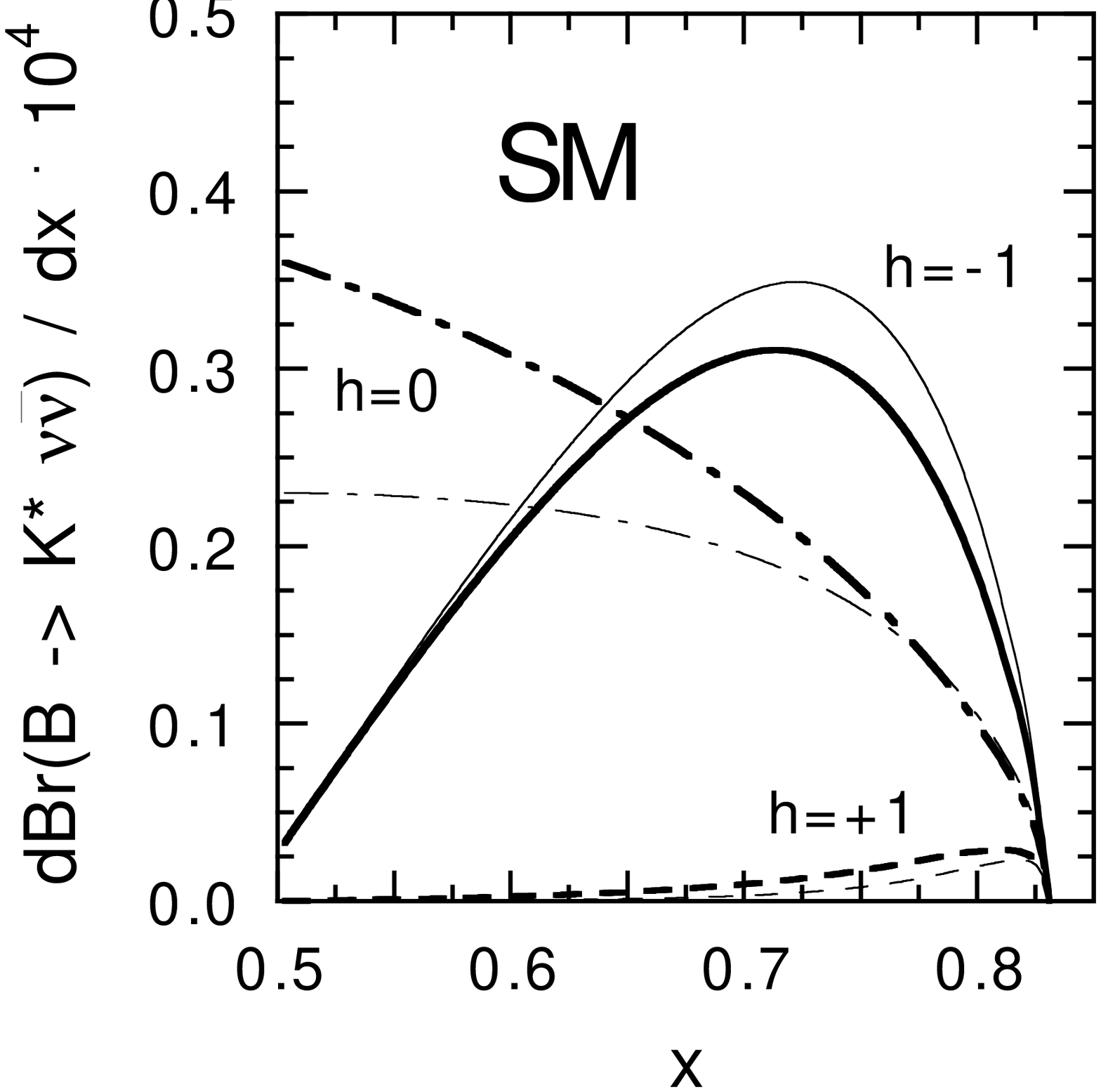}

\vskip -12.5cm

\parbox{10cm} \ $~~~~~~~~$ \ \parbox{6cm}{{\small \noindent
{\bf Figure 3}.  Differential branching ratio of the decay process 
$B \to K^* \nu \bar{\nu}$ (see Eqs. (\ref{BtoK*L}-\ref{BtoK*T}))
versus the missing-energy fraction $x$, calculated within the $SM$
framework. The dot-long dashed, dashed and solid lines correspond to
fina $K^*$-mesons with polarization $h = 0, +1, -1$, respectively.
The thin and thick lines have the same meaning as in Fig. 1.}}

\end{figure}

\vspace{2cm}

\begin{figure}[htb]

\epsfxsize=18cm \epsfig{file=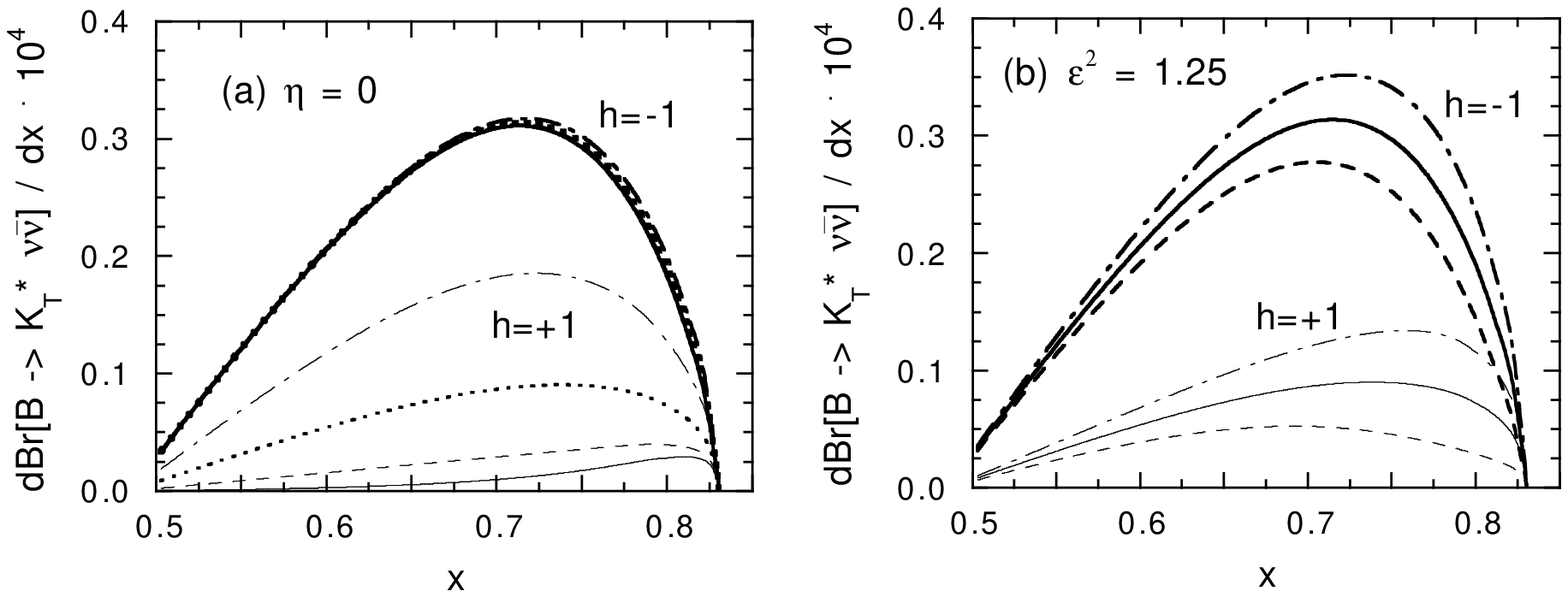}

\vspace{-18cm}

{\bf Figure 4}. {\small Missing-energy spectrum of transversely
polarized $K_T^*$-mesons produced in rare $B \to K^* \nu \bar{\nu}$ 
decays, calculated using our lattice-constrained dispersion quark model.
In (a) the parameter $\eta$ (Eq. (\ref{epseta})) is fixed at the value
$\eta = 0$, while the solid, dashed, dotted and dot-long dashed lines
correspond to $\epsilon^2 = 1.0, 1.05, 1.25$ and $1.55$, respectively.
The thin and thick lines are the results obtained in case of $RH$ and
$LH$ final $K_T^*$-mesons. In (b) the same as in (a), but at 
$\epsilon^2 = 1.25$ for various values of $\eta$; the dashed, solid 
and dot-long dashed lines correspond to $\eta = -0.25, 0.0$ and $0.25$, 
respectively.}

\end{figure}


\begin{thebibliography}{99}

\bibitem{BBMR91} See, e.g., S. Bertolini, F. Borzumati, A. Masiero and
 G. Ridolfi: Nucl. Phys. {\bf B353} (1991) 591.

\bibitem{IL81} T. Inami and C.S. Lim: Prog. Theor. Phys. {\bf 65} (1981)
 287.

\bibitem{BB93} G. Buchalla and A.J. Buras: Nucl. Phys. {\bf B400} (1993)
 225. G. Buchalla, A.J. Buras and M. Lautenbacher: Rev. Mod. Phys. {\bf
 65} (1996) 1125.

\bibitem{ALI96} For a recent review see, e.g., A. Ali: Acta Physica Pol.
 {\bf B27} (1996) 3529 and preprint DESY 97-192, September 1997 (e-print
 archive hep-ph/9709507).

\bibitem{GLN96} Y. Grossman, Z. Ligeti and E. Nardi: Nucl. Phys. {\bf
 B465} (1996) 369.

\bibitem{FLS96} A.F. Falk, M. Luke and M.J. Savage: Phys. Rev. {\bf D53}
 (1996) 2491.

\bibitem{MNS} D. Melikhov, N. Nikitin and S. Simula: preprint ISS-INFN
 97/15, e-print archive hep-ph/9711362.

\bibitem{COL97} P. Colangelo {\em et al.}: Phys. Lett. {\bf B395} (1997)
 339.

\bibitem{MEL96} D. Melikhov: Phys. Rev. {\bf D53} (1996) 2460; Phys. Lett.
 {\bf B380} (1996) 363; Phys. Lett. {\bf B394} (1997) 385; Phys. Rev. {\bf
 D56} (1997) 7089.

\bibitem{MNS97} D. Melikhov, N. Nikitin and S. Simula: Phys. Lett. {\bf
 B410} (1997) 290.

\bibitem{ISGW} D. Scora and N. Isgur: Phys. Rev. {\bf D52} (1995) 2783.

\bibitem{CAR94} F. Cardarelli {\em et al.}: Phys. Lett. {\bf B332} (1994)
 1.

\bibitem{GI85} S. Godfrey and N. Isgur: Phys. Rev. {\bf D32} (1985) 189.

\bibitem{SIM96}  I. L. Grach {\em et al.}: Phys. Lett. {\bf B385} (1996)
 317; Phys. Atom. Nucl. {\bf 59} (1996) 2152. S. Simula: Phys. Lett. {\bf
 B373} (1996) 193.

\bibitem{APE} $APE$ Collaboration, A. Abada {\em et al.}: Phys. Lett. {\bf
 B365} (1996) 275. 

\bibitem{UKQCD} $UKQCD$ Collaboration,  J. M. Flynn and C. T. Sachrajda,
 e-print archive hep-lat/9710057. J. M. Flynn {\it et al.}: Nucl. Phys.
 {\bf B461} (1996) 327.

\bibitem{LAT97} UKQCD Collaboration, L. Del Debbio {\it et al}, e-print
 archive hep-lat/9708008.

\bibitem{STE95} B. Stech: Phys. Lett. {\bf B354} (1995) 447; Z. Phys.
 {\bf C75} (1997) 245. 
 
\bibitem{LEL96} L. Lellouch: Nucl. Phys. {\bf B479} (1996) 353.

\bibitem{BHP96} T.E. Browder, K. Honscheid and D. Pedrini:  Ann. Rev.
 Nucl. Part. Sci. {\bf 46} (1996) 395.

\bibitem{PDG96} Particle Data Group: R.M. Barnet {\em et al}: Phys. Rev.
 {\bf D53} (1996) 1.

\bibitem{ALEPH} $ALEPH$ Collaboration: contribution PA10-019  to the Int.
 Conf. on {\em High Energy Physics}, Warsaw (Poland), 25-31 July, 1996 and
 D. Buskulic et al.: Phys. Lett. {\bf B343} (1995) 444.

\bibitem{RIZ98} T.G. Rizzo: preprint SLAC-PUB-7702, e-print archive
 hep-ph/9802401.

\bibitem{CLEO} $CLEO$ Collaboration, R. Ammar {\em et al.}: Phys. Rev.
 Lett. {\bf 71} (1993) 674 and CLEO CONF. 96-05 (1996).

\bibitem{ALIEV97} T. M. Aliev {\em et al.}: Phys. Rev. {\bf D56} (1997)
 4260; Phys. Lett. {\bf B400} (1997) 194.

\bibitem{IW} N. Isgur and M.B. Wise: Phys. Lett. {\bf B232} (1989) 113;
 Phys. Lett. {\bf B237} (1990) 527; Phys. Rev. {\bf D42} (1990) 2388. 

\end{thebibliography}
\end{document}